\documentclass[a4paper, oneside, twocolumn, notitlepage, 10pt]{extarticle_ecoc}
\usepackage{ecoc}

\begin{document}
\selectlanguage{english}    % Standard Language

%-------------------------------------------------- Title -----------------------------------------------------%

\title{Entanglement Purification by Integrated Silicon Photonics}%

%------------------------------------------------- Authors-----------------------------------------------------%

\author{
    Yonghe Yu\textsuperscript{(1)}, Siyan Zhou\textsuperscript{(2)},
    Mujtaba Zahidy\textsuperscript{(1)}, Caterina Vigliar\textsuperscript{(1)},
    Karsten Rottwitt\textsuperscript{(1)}, 
    \\Leif K. Oxenløwe\textsuperscript{(1)}, Yunhong Ding\textsuperscript{(1)}
}

\maketitle                  % Create title and author

%------------------------------------------ Description of Authors ----------------------------------------------%

\begin{strip}
    \begin{author_descr}

        \textsuperscript{(1)} Technical University of Denmark, Lyngby, Denmark, \textcolor{blue}{\uline{yudin@dtu.dk}}

        \textsuperscript{(2)} SiPhotonIC ApS, Virum, Denmark

    \end{author_descr}
\end{strip}

% \setstretch{1.1}
%-------------------------------------------------- Footnote -------------------------------------------------------%
\renewcommand\footnotemark{}
\renewcommand\footnoterule{}
%\let\thefootnote\relax\footnotetext{text}

%-------------------------------------------------- Abstract ---------------------------------------------------------%

\begin{strip}
    \begin{ecoc_abstract}
        % NOTE: Don't use a blank line here but start abstract right away to avoid an extra line break
        We demonstrate the first on-chip deterministic entanglement purification based on silicon photonics. To evaluate the purification performance, we simulate the bit-flip and phase-flip errors by reconfigurable circuits on chip. The state fidelity improves from 0.71 to 0.82 under a 20\% bit-flip error rate.
        \textcopyright2025 The Author(s)
    \end{ecoc_abstract}
\end{strip}

%-------------------------------------------------- Introduction Section -------------------------------------------------------%

\section{Introduction}
%远距离的量子纠缠是量子通信与分布式量子计算所需的最重要的基础资源之一，但纠缠的质量常常受到环境噪声与信道噪声的影响。由于能够提取高质量的纠缠态，量子纠缠纯化被认为是量子中继器的重要组成部分。自量子纠缠纯化提出以来，已有多个关于纠缠纯化理论与实验的报道。在这之中，借助超纠缠的确定性纠缠纯化消耗光子的纠缠自由度而不需要多对纠缠对参与，从而具有很高的纯化效率。
%为了实现可扩展的量子网络，量子协议被期望以更高的电子学与光学集成度实现。在这一趋势下，集成光子学因为高集成度、高扩展性、高稳定性与CMOS兼容，成为实现量子协议的重要平台。在量子通信协议中，量子纠缠交换与量子隐形传态已在集成光子学平台上实现。由于现有的纠缠纯化方案仍离不开块状光学晶体与光纤器件，片上的纠缠纯化实验仍然是一片空白。
Long-distance distributed quantum entanglement is one of the most important foundational resources for quantum communication and distributed quantum computing \cite{azuma2023quantum}. However, the entanglement fidelity can be degraded due to environmental noise. 
As a method to extract high-fidelity entangled states, quantum entanglement purification serves as a crucial component for quantum entanglement based applications\cite{puriandrepea1999,azuma2023quantum}. Since the first entanglement purification protocol was proposed \cite{puri41996}, entanglement purification has been studied theoretically and experimentally using bulk crystals \cite{puri12001,puri22003,puri32021}. For example, deterministic entanglement purification based on hyperentanglement consumes the entangled degrees of freedom of photons without requiring interaction between multiple entangled photon pairs, thereby achieving high purification efficiency\cite{puri32021}.

To realize scalable quantum networks, quantum protocols are expected to be implemented with stronger electronic and photonic integration\cite{pelucchi2022potential}. Silicon integrated photonics has emerged as a promising platform for implementing quantum protocols due to its high integration, scalability, stability, and compatibility with complementary metal-oxide-semiconductor (CMOS) technology\cite{pelucchi2022potential}. 
Quantum communication protocols such as entanglement swapping\cite{samara2021entanglement} and quantum teleportation\cite{llewellyn2020chipswapping} have already been demonstrated experimentally based on silicon photonics. However, existing entanglement purification schemes still rely on bulk crystals and fiber-based components, leaving on-chip entanglement purification experiments unexplored.

In this work, we experimentally demonstrate deterministic entanglement purification on a silicon-photonic chip. We designed and fabricated integrated circuits to generate path-entangled photon pairs, simulate the bit-flip (BF) error and phase-flip (PF) error in noisy fiber channels, and performed the purification and measurement. When a 20\% BF error rate is applied, the entanglement fidelity rises from 0.71 to 0.82 after purification, while the value of Clauser-Horne-Shimony-Holt (CHSH) inequality \cite{CHSH1} rises from 1.87 to 2.17. Our research proves the feasibility of using integrated photonic chips to quantum entanglement in noisy optical fibers.

\section{Experimental setup and results}
\begin{figure}[h]
    \centering
    \includegraphics[height=3.2cm]{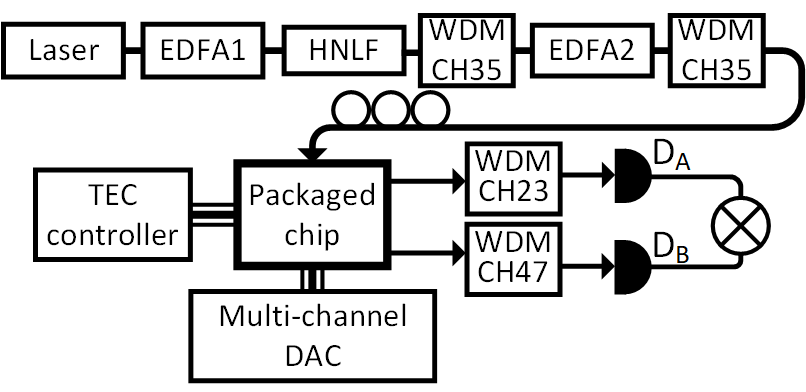}
    \caption{The experimental setup of the on-chip entanglement purification experiment. A pulsed laser with a 3 nm spectral bandwidth is utilized as pump light. To broaden the spectrum and ensure coverage at 1549.32 nm, we amplified the laser and passed it through a highly nonlinear fiber (HNLF). The chip is packaged with electrical wires bonding, optical fiber arrays, a thermistor, and a thermoelectric cooler (TEC). The detection efficiency of the SPDs is 25\%. }
    \label{fig1}
\end{figure}
The experimental setup is depicted in Fig.~\ref{fig1}. %The pump laser is a 10 GHz pulsed laser (measured as 9.97 GHz) and amplified by a pre-amplifier (EDFA1). The light is then filtered by a WDM filter with 200 GHz bandwidth and 1549.32 nm center wavelength. After one more 1549.32 nm WDM filter and a polarization controller, the optical power is measured as 19.2 dBm. 
The pump laser is a 10 GHz pulsed laser (measured as 9.95 GHz), which is first amplified by a pre-amplifier (EDFA1). The amplified pulses then pass through a highly nonlinear fiber (HNLF), followed by a wavelength-division multiplexing (WDM) filter with a 200 GHz bandwidth centered at 1549.32 nm. The filtered pulses are subsequently amplified by a high-power amplifier (EDFA2). After passing through the second WDM filter and a polarization controller, the optical power is measured to be 19.2 dBm. The filtered laser is coupled into the chip via a single-mode fiber array (FA). The signal and idler photons generated through spontaneous four-wave mixing (SFWM) in the waveguide are also coupled out using a single-mode FA. To get rid of the pump photons and ensure the spectral purity, the signal and idler photons are filtered with two 100 GHz bandwidth WDM filters at 1539.77 nm and 1558.98 nm, respectively. The filtered photons are then detected by two single-photon detectors (SPD), and the coincidence is recorded by a time tagger.

\begin{figure*}[t]
    \centering
    \includegraphics[height=4.3cm]{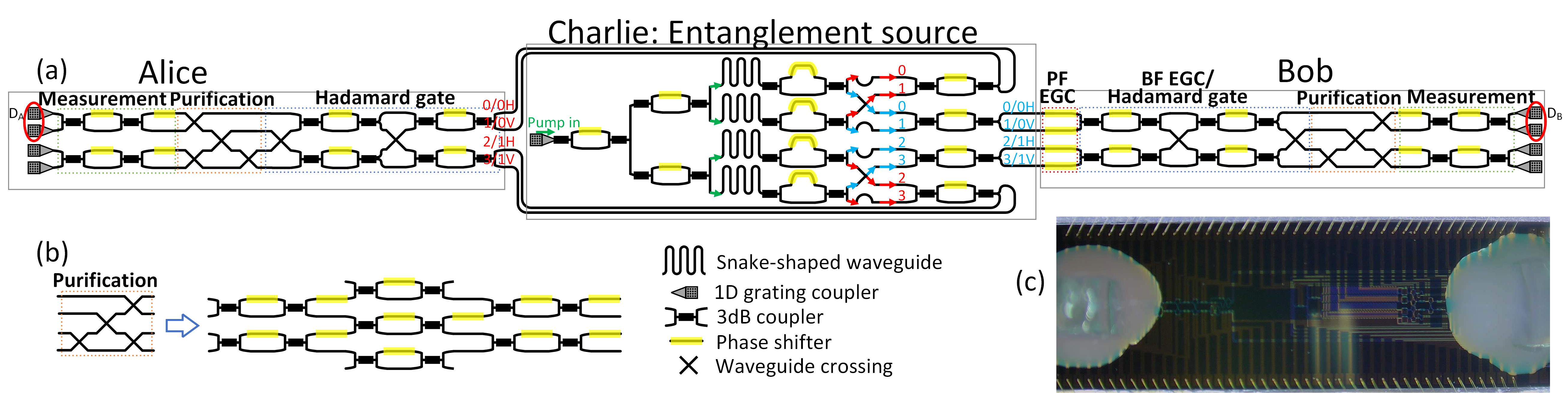}
    \caption{The circuit schematic of the entanglement purification chip: (a) The schematic includes three parts: Charlie for entangled photon generation, Alice and Bob for purification demonstration. The coupling efficiency of the grating coupler here is measured as -5.4 dB at 1550 nm, and the waveguide loss is around 4.5 dB/cm for the TE mode. (b) The reconfigurable purification circuit. (c) The packaged silicon chip to demonstrate entanglement purification.}
    \label{fig2}
\end{figure*}

As shown in Fig.~\ref{fig2}(c), the chip is packaged with electrical wires mounted on a printed circuit board (PCB), allowing it to be controlled by a multi-channel digital-to-analog converter (DAC). The pump is coupled into the circuit as shown in Fig.~\ref{fig2}(a). The circuit in this work consists of three parts: Charlie for entanglement generation, Alice for purification and measurement, and Bob for noise simulation, purification and measurement. In the Charlie part, the pump power is equally distributed into four 1.5 cm long waveguides, where SFWM occurs and generates one pair of signal and idler photons. The initially generated path-encoded entangled state is given by:$\frac{1}{2}\left(\left|00\right\rangle + \left|11\right\rangle + \left|22\right\rangle + \left|33\right\rangle\right)$. %When the SPDs receive the photon pair from the first pair of waveguides, the count rate of both channels is approximately 70 kHz, and the coincidence rate is around 40 Hz.
The photon pair coincidence rate is around 40 Hz, limited by the packaging loss.

\begin{figure}[h]
    \centering
    \includegraphics[height=5.4cm]{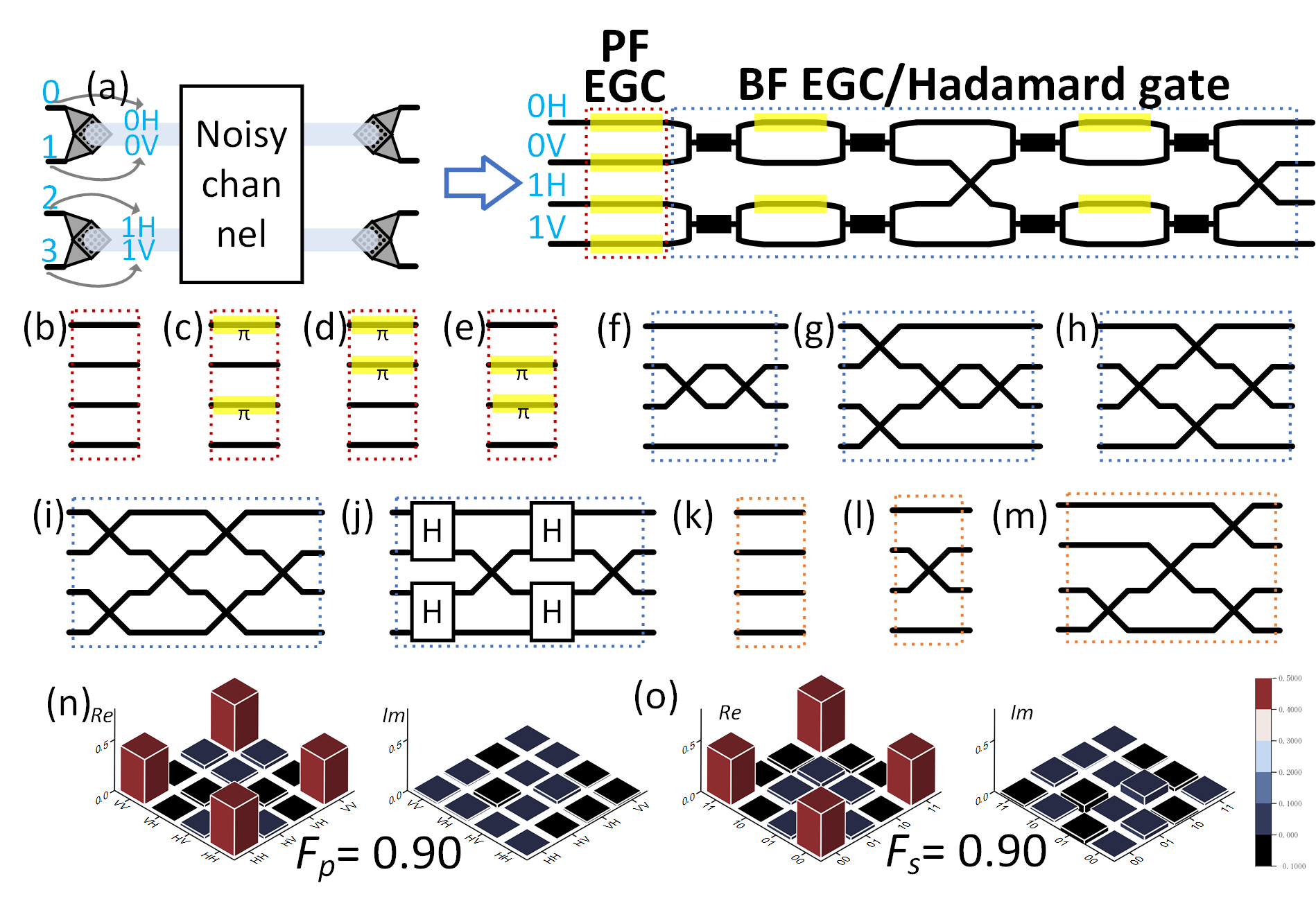}
    \caption{Schematic of the method we use to simulate the fiber error. (a) Coupling from four on-chip waveguides to two optical fibers using two 2D grating couplers. (b)\~{}(e) The PF error generation circuit when no error (b), PF error on polarization qubit (c), PF error on spatial qubit (d), and PF error on both polarization and spatial qubit (e). (f)\~{}(i) The BF error generation circuit when no error (f), BF error on polarization qubit (g), BF error on spatial qubit (h), and BF error on both qubits (i). (j) Implementation of Hadamard gates to convert the PF error to the BF error. (k)\~{}(m) The purification circuit when purification off and polarization tomography (k), purification off and spatial tomography (l), and purification on and polarization tomography (m). (n) The result for polarization tomography when no error. (o) The result for spatial tomography when no error.}
    \label{fig3}
\end{figure}

The chip we designed in this work is aimed at purifying the entanglement in noisy fiber channels. %In Fig.~\ref{fig3}a, we list a typical way to achieve the chip-fiber coupling by the 2D grating coupler. In this way, the photon in the first waveguide is converted into the horizontal photon in the first fiber ($\left|0\right\rangle\rightarrow\left|0H\right\rangle$), and 
In this work, we use the path entangled state to represent the hyper-entanglement in the fiber instead of sending the photons into the real fiber. Fig.~\ref{fig3}(a) illustrates a scenario where our purification scheme can be applied. With 2D grating couplers, the waveguide mode $\left|0\right\rangle$ could be converted into fiber-based $\left|0H\right\rangle$, and $\left|1\right\rangle$ could be converted into fiber-based $\left|0V\right\rangle$.
Therefore, the initial path-encoded entangled state is equal to $\left|\psi_0\right\rangle=\frac{1}{2}\left(\left|0H0H\right\rangle + \left|0V0V\right\rangle + \left|1H1H\right\rangle + \left|1V1V\right\rangle\right)=\frac{1}{2}\left(\left|00\right\rangle + \left|11\right\rangle \right)\left(\left|HH\right\rangle + \left|VV\right\rangle \right)$. With the measurement circuits in Alice and Bob, we performed quantum state tomography \cite{llewellyn2020chipswapping}, and the fidelity of the polarization qubit and spatial qubit is 0.90 and 0.90 respectively, which are limited by dark counts, the extinction ratio of the Mach–Zehnder interferometer (MZI), and imperfect heater calibration.

\begin{figure*}[t]
    \centering
    \includegraphics[height=3.8cm]{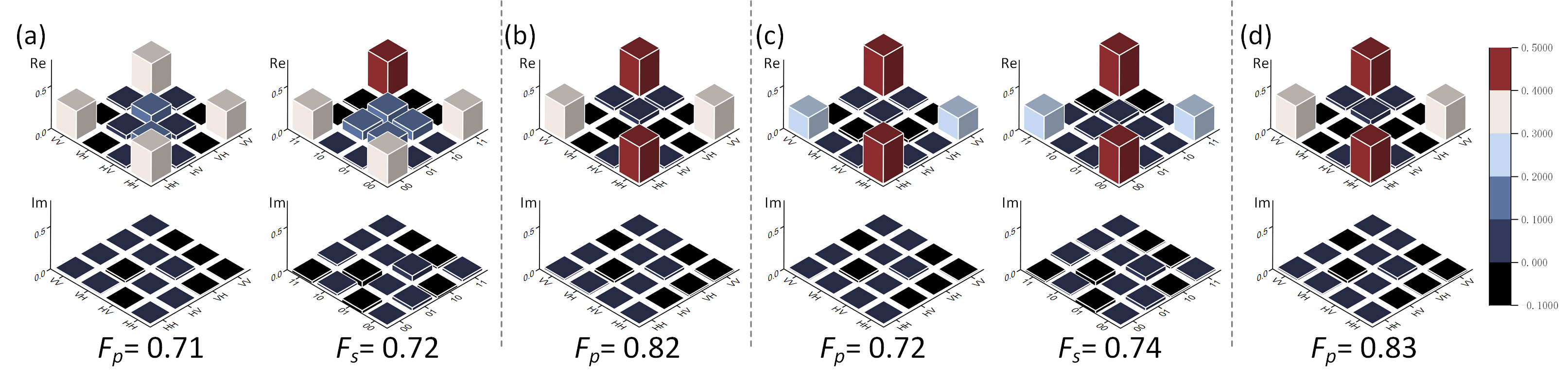}
    \caption{The tomography results before and after entanglement purification. With 20\% BF error, the density matrix of polarization qubit and spatial qubit before purification (a) and the density matrix of polarization qubit after purification (b). With 20\% PF error, the density matrix of polarization qubit and spatial qubit before purification (c) and the density matrix of polarization qubit after purification (d).}
    \label{fig4}
\end{figure*}

To simulate the BF error in noisy fiber channels, a BF error generation circuit (EGC) at the Bob's part is designed and reconfigured to the circuits as shown in Fig.~\ref{fig3}(f)-\ref{fig3}(i) in a time distribution to satisfy a 20\% bit-flip rate in the polarization and spatial dimension \cite{puri32021}. Taking the circuit in Fig.~\ref{fig3}g as an example, the state is changed from $\left|\psi_0\right\rangle$ to $\frac{1}{2}\left(\left|00\right\rangle + \left|11\right\rangle \right)\left(\left|HV\right\rangle + \left|VH\right\rangle \right)$, which corresponds to the BF noise being applied on the polarization qubit. In the case that the EGC is configured to the circuit as shown in Fig.~\ref{fig4}(h), the BF noise is applied on the spatial qubit. When a 20\% overall bit-flip rate is applied, the fidelity of the polarization qubit is characterized as 0.71 with respect to the ideal Bell state $\frac{1}{\sqrt{2}}\left(\left|HH\right\rangle + \left|VV\right\rangle \right)$, as shown in Fig.~\ref{fig4}(a). With an extra circuit as shown in Fig.~\ref{fig3}(i), the fidelity of the spatial qubit is characterized as 0.72 with respect to the Bell state $\frac{1}{\sqrt{2}}\left(\left|00\right\rangle + \left|11\right\rangle \right)$.

In order to simulate the PF error before purification, the MZIs of the EGC are configured as identity, and different phase configurations are applied leading to identical circuits of Fig.~\ref{fig3}(b)-\ref{fig3}(e). Distributing these circuits in time  \cite{puri32021} leads to efficient simulation of 
%As illustrated in Fig.~\ref{fig3}b\~{}e, the circuit presented in this work enables the application of 
PF noise on either the polarization or spatial qubit. With a 20\% PF rate, the fidelity of the polarization qubit and spatial qubit is 0.72 and 0.74 with respect to $\frac{1}{\sqrt{2}}\left(\left|HH\right\rangle + \left|VV\right\rangle \right)$ and $\frac{1}{\sqrt{2}}\left(\left|00\right\rangle + \left|11\right\rangle \right)$, respectively.

The reconfigurable purification circuit shown in Fig.~\ref{fig2}(b) can be activated at both the Alice and Bob sides. If the purification is activated, the purification circuits are configured into the configuration shown in Fig.~\ref{fig3}(m), which equivalently changes the quantum state $\left|0H\right\rangle\rightarrow\left|0V\right\rangle$, $\left|0V\right\rangle\rightarrow\left|1V\right\rangle$, $\left|1H\right\rangle\rightarrow\left|1H\right\rangle$, and $\left|1V\right\rangle\rightarrow\left|0H\right\rangle$. The spatial qubit is then discarded, and only the outputs from the first two waveguides are selected to collect the purified entangled state. In this way, the state with one BF error on either polarization qubit or spatial qubit results in no coincidence. Since the probability of BF errors occurring on both polarization and spatial dimensions simultaneously is much lower than BF error on a single dimension, this scheme can significantly improve the entanglement fidelity. We take the error state $\frac{1}{2}\left(\left|00\right\rangle + \left|11\right\rangle \right)\left(\left|HV\right\rangle + \left|VH\right\rangle \right)$ as an example: after the purification circuit, the state is $\frac{1}{2}\left(\left|0V1V\right\rangle + \left|1V0V\right\rangle + \left|1H0H\right\rangle + \left|0H1H\right\rangle \right)=\frac{1}{2}\left(\left|13\right\rangle + \left|31\right\rangle + \left|20\right\rangle + \left|02\right\rangle\right)$. The photons from the first two pairs of GCs (red circle in Fig.~\ref{fig1}(a)) are recorded to check the coincidence, and thus, the states with one BF error are excluded from the experimental dataset. In our experiment on purification after BF error, as shown in Fig.~\ref{fig4}(b), the fidelity of the purified entangled state increases from 0.71 to 0.82, representing a more than 10\% improvement compared to the fidelity before purification.
In the BF error regime, the value of the CHSH inequality, measured by reconfiguring the MZIs in the measurement circuit, increases from 1.87 to 2.17, which breaks the CHSH inequality.

Since the purification circuit in this work only processes the BF error, we use the Hadamard gate to convert the PF error to the BF error. Although the EGC is only activated at Bob's side, the Hadamard gate needs to be applied to both Alice and Bob.
Therefore, in addition to reconfiguring the EGC of Bob's side into four Hadamard gates, the four MZIs on Alice's side are also reconfigured as Hadamard gates (Fig.~\ref{fig4}(j)). After applying the Hadamard gates and performing entanglement purification, the fidelity of the entangled state increases from 0.72 to 0.83, and the value of the CHSH inequality increases from 1.94 to 2.19.

The purification circuit we propose here, essentially, incorporates the functionality of the controlled NOT (CNOT) gate. In the purification protocol presented in this work, what we actually consume is the degrees of freedom of photons rather than the number of photon pairs. %As a result, the fidelity of entangled photon pairs collected from waveguides 2 and waveguide 3 could also be improved.
As a result, the purified photon pairs could also be collected from another two pairs of GCs in Fig.~\ref{fig1}(a).
%-------------------------------------------------- Conclusions Section ———————————————————————————%

\section{Conclusions}
We experimentally demonstrate entanglement purification on a silicon chip for the first time. The purification protocol employed here is based on a deterministic scheme, resulting in high purification efficiency \cite{puri32021}. By using 2D grating couplers, our entanglement source circuit can generate polarization-spatial hyper-entangled photons in fiber, and the purification circuit can purify the BF error and PF error. Since noise in optical fibers can be decomposed into a combination of BF error and PF error, our circuit can effectively improve the fidelity of entanglement in quantum communication over noisy channels. 

%-------------------------------------------------- Acknowledgements Section -------------------------------------------------------%
%\clearpage
%\section{Acknowledgements}
%One extra page is allowed so that acknowledgements and references can be given in full length.

%-------------------------------------------------- Bibliography Section -------------------------------------------------------%
% see also https://tex.stackexchange.com/questions/55030/text-before-references-but-after-bibliography-title-with-bibtex as of 2024-02-29
\defbibnote{myprenote}{
}
\printbibliography[prenote=myprenote]

\vspace{-4mm}

%%%%%%%%%%%%%%%%%%%%%%%%%%%%%%%%%%%%%%%%%%%%%
%---------------------------------------------- End of Document -----------------------------------------------%
\end{document}